\documentclass[12pt,preprint]{aastex}

\begin{document}

\title {WHAT NEXT FOR THE LIKELY PRE-SUPERNOVA, HD 179821?}

\author{M. Jura}
\affil{Department of Physics and Astronomy, University of California,
    Los Angeles CA 90095-1562; jura@clotho.astro.ucla.edu}

\author{T. Velusamy and M. W. Werner}
\affil{Jet Propulsion Laboratory, 169-506, 4800 Oak Grove Dr., Pasadena CA 91109; velu@rams.jpl.nasa.gov; mwerner@sirtfweb.jpl.nasa.gov}

\begin{abstract}
We have used the Owens Valley Radio Observatory  Millimeter Array to obtain a map of the J = (1${\rightarrow}$0) CO emission from the circumstellar shell around HD 179821, a highly evolved G-type star which will probably explode as a supernova in the next 10$^{5}$ yr.  Very approximately, the gas presents as a circular ring with
an inner diameter of 3${\farcs}95$, an outer diameter of 
${\sim}$12${\arcsec}$ and with
azimuthal variations in the CO brightness by about a factor of 2.   Until about 1600 years ago, the star was a red hypergiant losing about 3 ${\times}$ 10$^{-4}$ M$_{\odot}$ yr$^{-1}$ at an average outflow speed of 32 km s$^{-1}$.  We propose that when HD 179821 explodes as a supernova, it  may resemble Kepler's supernova remnant and thus some of the anisotropies in supernova remnants may be intrinsic.   If the factors which cause the anisotropic mass loss in HD 179821 persist
 to the moment when the star explodes as 
a supernova, the  newly-born pulsar may receive a momentum ``kick" leading to a space motion near ${\sim}$700 km s$^{-1}$.  Independent of the angular asymmetries, the radially detached  shell around HD 179821 may be representative of environments which produce dust echoes from gamma-ray bursts.  
\end{abstract}

\section{INTRODUCTION}
 After main sequence O-type stars of more than 20 M$_{\odot}$ with $T_{eff}$ $>$ 30,000 K 
 consume their interior hydrogen, they evolve to become red hypergiants with  mass loss rates near 10$^{-4}$ M$_{\odot}$ yr$^{-1}$ (see, for example, Chin \& Stothers 1990, Schaller et al. 1992).  Ultimately, the star explodes as a type II supernova.  Because deviations from spherical symmetry might be crucial in  supernova explosions (Burrows 2000), we hope to gain insights into the physics of  supernovae by studying the anisotropic circumstellar shells of pre-supernovae.  

The morphology of
a supernova remnant is determined both by the history of mass
loss from the pre-supernova and the structure of the supernova explosion itself.    We hope to learn whether nonspherical shapes in supernova remnants result from  nonspherical mass loss by the pre-supernova
 star, from  intrinsically nonspherical
supernova explosions, from irregularities in the surrounding interstellar
 medium, or from some combination of these different effects.

A mystery related to supernova explosions is that the newly-born pulsars sometimes have space motions over 700 km s$^{-1}$ (Cordes \& Chernoff 1998).  It is possible that these motions result from a ``momentum-kick" acquired during a non-spherical explosion,  another reason for studying asymmetries in the outflows from pre-supernovae.

Gamma-ray bursts may be produced during supernova explosions.  If so, then
dust at 0.1 to 1 pc from the star may result in a ``dust echo" some 20-30 days after the initial burst (Esin \& Blandford 2000).  The morphology of dust around massive stars
may provide constraints on models for dust echoes.

The winds from red hypergiants typically contain large amounts of dust, and therefore these objects are strong infrared sources (see, for example, Jura \& Kleinmann 1990).  
As discussed both by Jura \& Werner (1999) and below, it is likely, though controversial, that
HD 179821 (m$_{V}$ = 7.9 mag, b = -4.96$^{\circ}$, spectral type G5Ia, and a strong IRAS source with $F_{\nu}$(25 ${\mu}$m) = 650 Jy) is a highly evolved  star near the Humphreys-Davidson limit with an initial mass near 30 M$_{\odot}$ that will become a supernova within the next ${\sim}$10$^{5}$ yr. The alternative model is that it was a
${\sim}$1 solar mass star that has become a pre-Planetary Nebula, and it lies at a distance of about 1 kpc from the Sun.    Although dust nebula  are observed around post main sequence hypergiants  with luminosities greater than 10$^{5}$ L$_{\odot}$ 
(de Jager 1998, Nota et al. 1995, Voors et al. 2000), most of these stars lie in the Galactic Plane where interactions with the local interstellar medium
are important.  Because HD 179821 is removed from the Galactic Plane, it may be a circumstellar shell with relatively little contamination by
swept-up interstellar gas.  About ${\sim}$20\% of O-type stars are runaways (Blaauw 1993) with speeds greater than 50 km s$^{-1}$, and these
stars can travel as much as 1000 pc from the Galactic Plane before they explode.
We propose that HD 179821 is post-main sequence, runaway, massive star.  
 
Previously, we have reported Keck observations of the 11.7 ${\mu}$m emission from HD 179821 and found that the circumstellar shell is very approximately a circular ring with an inner diameter of about 3${\farcs}$5 (Jura \& Werner 1999). However, both
the 11.7 ${\mu}$m emission and the map of the OH maser spots in the circumstellar envelope of HD 179821 (Claussen 1993) display significant deviations from circular symmetry.   
A major difficulty with interpreting the 11.7 ${\mu}$m map is that at this wavelength, $h{\nu}/k$ = 1230 K, while the grains are typically colder than 140 K.  As a result, the emission is from the Wien portion of the Planck curve and small uncertainties in the temperature translate into large uncertainties
in estimating the mass.    As with most bright evolved stellar infrared sources, HD 179821 also is a strong CO source (Zuckerman \& Dyck 1986, Bujarrabal, Alcolea \& Planesas 1992, van der Veen et al. 1993, Knapp et al. 2000, Josselin \& Lebre 2001). Therefore, in order to achieve
a better tracer of the mass loss history and  kinematics of the outflow, we have acquired a high resolution map of the circumstellar CO emission with the Owens Valley Radio Observatory Millimeter Array. 

In order to learn more about the evolution of supernovae and their remnants,  we  compare
our maps of the shell around HD 179821 with the appearance of Kepler's supernova remnant which, with an age of 400 years, is one of the youngest known remnants in the Milky 
Way (Burrows 2000). Since it lies at b = 6.8$^{\circ}$ and is at a distance from the Sun of ${\sim}$ 4800 pc (Reynoso \& Goss 1999), it is ${\sim}$500 pc from the Galactic Plane where there is relatively little interstellar matter.  Kepler's remnant thus presents an excellent opportunity to study the interaction of a supernova explosion with its pre-supernova circumstellar shell.  Kepler's supernova remnant probably (Decourchelle \& Ballet 1994, Rothenflug et al. 1994, Hughes 1999)  though not certainly (Kinugasa \& Tsunemi 1999) resulted from the explosion of a massive star.  

In Section 2, we summarize our picture of HD 179821 and present an additional argument for why we think that it is a massive star.  In Section 3, we present our  observations.  In section 4, we present a simple model to explain the data while in Section 5 we compare the asymmetries in the circumstellar matter around HD 179821 with those in 
Kepler's supernova remnant.  In Section 6, we note how our data may help both understand why pulsars may have space motions near 700 km s$^{-1}$ and the dust echoes from gamma-ray bursts.  We present our conclusions in Section 7.

\section{THE DISTANCE AND LUMINOSITY OF HD 179821}
The distance to HD 179821 is not well determined by trigonometric parallax (the value measured with the {\em Hipparcos} satellite was 0.18 ${\pm}$ 1.12 ${\times}$ 10$^{-3}$ {\arcsec}), and indirect means must be used.  Two models have been proposed.  (1) The star might be  at a ``large" distance of about 6000 pc, have a luminosity near 3 ${\times}$ 10$^{5}$ L$_{\odot}$ and thus have had an initial main sequence mass near 30 M$_{\odot}$ (Hawkins et al. 1995).  (2) The star might lie at a ``small" distance of about 1 kpc, have a luminosity near 10$^{4}$ L$_{\odot}$ and be the post-Asymptotic Giant Branch (AGB), pre-Planetary Nebula descendant of a star that was initially ${\sim}$1 M$_{\odot}$ on the main sequence (Hrivnak, Kwok \& Volk 1989). While   inferences of the luminosity of the star from the atmospheric abundances might resolve this argument; to date they are controversial (Reddy \& Hrivnak 1999, Lobel \& Dupree 2000, Thevenin, Parthasarathy \& Jasniewicz 2000), and not evidently conclusive.  
The detection of CO and Na I emission at 2.2 ${\mu}$m  (Hrivnak, Kwok \& Geballe 1994, Oudmaijer et al. 1995) suggests that standard model atmospheres may not apply, and 
additional methods to estimate the distance should be investigated.

Josselin \& Lebre (2001) report a well calibrated measure of the ratio of 
integrated intensities in the J = (2${\rightarrow}$1) lines of $^{12}$CO and $^{13}$CO of 3.2 from which they argue that HD 179821 is a post-AGB star.  However, Bujarrabal et al. (1992) reported a value of 9.6 for this ratio, and 6.6 for the ratio
of the  integrated intensities of the $^{12}$CO and $^{13}$CO J = (1${\rightarrow}$0)
lines.  Given that the hypergiant IRC+10420 has a value of $^{12}$CO/$^{13}$CO of
9 ${\pm}$ 2 (Fix \& Cobb 1987) and that the CO rotational lines are optically thick, observations of the
circumstellar CO isotope ratio do not  provide a compelling method for determining the luminosity and origin of the star. 

 From the properties of the circumstellar matter, we argue that HD 179821 is a massive, distant star.  
  Zuckerman \& Dyck (1986) first noted that the high LSR velocity of the star (+100 km s$^{-1}$) and high average outflow velocity of the gas (32 km s$^{-1}$) are best understood if the star lies at ${\sim}$6 kpc.  
There is a rough correlation between the outflow velocity, $V_{\infty}$, from a red giant
and its luminosity (Jones, Hyland \& Gatley 1983). Therefore, since $V_{\infty}$ for HD 179821 is near 32 km s$^{-1}$ (see below), which is significantly larger than
the typical value for a
 mass-losing red giants of 15 km s$^{-1}$, it is plausible that this star has a relatively high luminosity.  Barnbaum, Kastner \& Zuckerman (1991) found in a sample of 124   AGB carbon stars that 4 stars  have wind outflow speeds greater than 30 km s$^{-1}$.  These 4 objects lie at low galactic latitudes and presumably have main sequence progenitor masses over 2.5 M$_{\odot}$. If
HD 179821 lies at a distance of 1 kpc, then its high LSR radial velocity  indicates that it is a member of an old population with a main sequence progenitor mass less than 1 M$_{\odot}$, an inference not consistent with its relatively high outflow velocity.     

In addition to comparing $V_{\infty}$ of  HD 179821 with mass-losing red giants, we can also compare its outflow velocity with the measured values for Planetary Nebula.   
 Approximately 10\% of all Planetary Nebulae have expansion speeds greater than 30 km s$^{-1}$ (Sabbadin 1984, Weinberger 1989).  However,  such speeds appear to be a consequence of acceleration of the wind material after the 
star evolves beyond being a mass-losing red giant, since Gesicki \& Zijlstra (2000)
 find that expansion speeds greater than 30 km s$^{-1}$
are usually found only for Planetary Nebulae with radii larger than 0.03 pc.  For a presumed distance from the Sun of 1 kpc, the  current  radius of the peak
density around HD 179821 is 0.008 pc.   We conclude that if HD 179821 is a pre-Planetary Nebula with  an outflow speed greater than 30 km s$^{-1}$ and the densest material lying within 0.008 pc of the star, then it is very unusual and perhaps unique.      

The relatively
high value of the LSR velocity for HD 179821 is more easily understood if it is a distant object and its speed 
is a result of Galactic rotation. 
 Figure 1 shows a plot of LSR velocity vs. CO outflow velocity for  a complete sample  of high mass-loss
 rate (${\geq}$ 10$^{-6}$ M$_{\odot}$ yr$^{-1}$) oxygen-rich AGB stars that are within 1 kpc of the Sun and further north than ${\delta}$ = -32$^{\circ}$ (Jura \& Kleinmann 1989).  We also show on this plot the location of HD 179821 ($l$ = 35.6, $b$ = -5.0) and the well known hypergiant IRC +10420 ($l$ = 47.0, $b$ = -2.5, $D$ = 5 kpc, $L$ = 5 ${\times}$ 10$^{5}$ L$_{\odot}$, Jones et al. 1993) which lies relatively close to HD 179821 in the sky.   The better agreement of the kinematics of the outflow from HD 179821 with those from  IRC+10420,
rather than with the outflow kinematics of the local AGB stars suggests that HD 179821 is a red hypergiant.
If
HD 179821 is a massive, runaway star, then its radial velocity cannot serve as
 an accurate measure of its distance.  However, if it does lie at 6 kpc from the Sun, then at its Galactic longitude, it lies beneath the 5 kpc ring, the region in the disk of the Milky Way with the greatest concentration of interstellar CO and giant H II regions (Scoville \& Sanders 1987).  Therefore, if HD 179821 is a hypergiant, a distance of 6 kpc from the Sun is plausible but not certain.  

Another argument for the high luminosity of 
 HD 179821 is that it has a near-infrared reflection nebulae similar to that of the hypergiant IRC+10420 (Kastner \& Weintraub 1995).  
Furthermore,  Jura \& Werner (1999) point out that the OH maser map presented by Claussen (1993) is more naturally understood if the star lies at a distance of 6000 pc rather than 1000 pc. That is, if HD 179821 lies at ``only" 1000 pc from the Sun, then a uniquely high ultraviolet circumstellar opacity is required to account for  the  spatial extent of the OH maser emission which is determined by the depth of penetration of
the ambient interstellar radiation field
into the outer regions of the circumstellar shell.    

Finally, based on its kinematics and historic data, we present another argument  that HD 179821 probably lies further than 1 kpc from the Sun and thus probably has a distance near 6000 pc.  
The average expansion speed of 32 km s$^{-1}$ of the circumstellar matter is determined from the profile of the circumstellar CO emission  and is therefore independent of distance.  In contrast, the inferred physical diameter of the inner boundary of the circumstellar is derived from both the measured inner angular diameter of 3${\farcs}$5 and the assumed distance.  
If HD 179821 lies at 6000 pc from the Sun,  the time required for the gas in  the inner shell to expand to its current size is 1600 yr.  Therefore, there would not have been much
change in the star during the past 100 yr.  On the other hand, if the star has just left the AGB and
lies at a distance of 1000 pc, then the time to expand to its current size
is only 260 yr, and during the past 100 yr, it could have undergone observable changes.  Such historic  changes of the light from post-AGB stars in fact have been measured.  For example, during  the past 100 years, the Egg Nebula, a famous post-AGB star, has brightened  from m$_{B}$= 15 mag to m$_{B}$ = 13 mag (Gottlieb \& Liller 1976).  However, there is no historic evidence for any major changes in either the brightness or spectral type of HD 179821.   In the Henry Draper catalog, 
which is nearly 100 years old, the spectral type is G5, the same spectral type given in the SIMBAD data base from modern observations.    Since the change in spectral type from G5 either to G8 or to G2 corresponds to a change in the effective temperature of the star of ${\sim}$250 K (Drilling \& Landolt 2000), then it
appears that during the last 100 years, $dT_{eff}/dt$ ${\leq}$ 3 K yr$^{-1}$
 for HD 179821. In order to
have evolved from the AGB when it was losing a large amount of mass with an effective temperature of 2600 K to
its current effective temperature of either 6750 K (Reddy \& Hrivnak 1999) or 5660 K (Thevenin et al. 2000) within 260 years, then the contraction of the star and consequent rise in its surface temperature should have proceeded with an
average value of $dT_{eff}/dt$ of ${\geq}$11 K yr$^{-1}$,  which is larger than inferred.  Furthermore, the magnitude of 8.1 given in the ${\it Bonner Durchmusterung}$ catalog, which was compiled around 1860, is approximately the same as its current average visual magnitude of 7.9; there is no evidence for major changes during the past 140 years.  Therefore,   the historic evidence supports the
view that HD 179821 is a distant, luminous star.       

\section{OBSERVATIONS OF THE CIRCUMSTELLAR CO}
We used the six-element Owens Valley Radio Observatory Millimeter Array during November - December 1999 to observe the J = (1${\rightarrow}$0) $^{12}$CO emission from HD 179821.  Phase calibration was performed on the quasar 1749+096 while flux
calibrations were derived from 3C 273 and 3C 454.3.  The synthesized beam
was 2{\farcs}0 ${\times}$ 1{\farcs}35 at position angle -85$^{\circ}$;  the 1${\sigma}$ rms noise was 40 mJy/beam.  
The 
map of the integrated CO intensity is shown in Figure 2 where  the contour levels
are spaced by 1.8${\sigma}$ with each level corresponding to 1.07 K or
32.5 mJy beam$^{-1}$.  The channel maps with 
5.35 km s$^{-1}$ resolution are shown in Figure 3.  The channel map velocities  are given relative to the assumed LSR velocity of  100 km s$^{-1}$
(Bujarrabal et al. 1992).  The line profiles are not symmetric, and
estimates of line center have ranged 95 and 105 km  s$^{-1}$ (Reddy \& Hrivnak 1999;  Josselin \& Lebre 2001).  The position of the star was taken from the Hipparcos data as ${\alpha}$(2000.0) = 19$^{h}$ 13$^{m}$ 58.61$^{s}$, ${\delta}$(2000.0) = 00$^{\circ}$ 07${\arcmin}$ 31${\farcs}$93 and is shown on our maps.

We show in Figure 4 the azimuthally-averaged intensity as a function of angular offset from the star in the channel at line center and at a velocity offset of +27 km s$^{-1}$.  As can be seen from Figure 4, the peak of the CO emission is seen to have an angular diameter of 3${\farcs}$95,  slightly larger than the 3${\farcs}$5 diameter  found in the 11.7 ${\mu}$m image (Jura \& Werner 1999). In our CO data, the peak to the north is about a factor of 1.4 times brighter than the peak to the south; in the 11.7 ${\mu}$m image, the contrast between the northern and southern peaks is about a factor of 2 (Jura \& Werner 1999).  In the east-west direction, the shell is less well defined.   As shown in Figure 4, the CO emission can be traced to a radius of  about 6${\arcsec}$ so that the diameter of the CO emission is at least 12${\arcsec}$.  The total flux measured with the interferometer in this
12${\arcsec}$ region is 45\% - 55\% of the flux measured with the 30m IRAM telescope with a beam diameter of 22${\arcsec}$ (Bujarrabal et al. 1992, Josselin \& Lebre 2001).  Therefore, the CO is  somewhat more extended than shown in our data.  Figure 5 shows the CO spectrum at the center of
the circumstellar envelope.

\section{CIRCUMSTELLAR GAS DISTRIBUTION AROUND HD 179821}

\subsection{A first approximation: spherical geometry}

In order to estimate the mass loss rate when the star was a red giant, we start with the simplification of spherical symmetry.
An estimate of the gas loss rate during the red hypergiant phase, ${\dot M}$ (M$_{\odot}$ yr$^{-1}$), can be derived from the single-beam measurements of the circumstellar CO emission.  According to Kastner (1992), for observations of the $J$ = (1${\rightarrow}$0) transition of CO with the 30m IRAM telescope, if ${\dot M}$ $<$ 10$^{-5}$ M$_{\odot}$ yr$^{-1}$, then
\begin{equation}
{\dot M}\;=\;6\,{\times}\,10^{-9}\;T_{mb}(0)^{0.75}\;V_{\infty}^{2}\,D^{2}
\end{equation}
where $T_{mb}(0)$ is the main beam antenna temperature (K) at line center,  $V_{\infty}$ is the gas outflow velocity (km s$^{-1}$),  and $D$ is the distance (kpc) of the star from the Sun.   In this expression, we assume that in the circumstellar envelope, [CO]/[H$_{2}$] = 10$^{-3}$,
consistent with  the abundance analysis of the photosphere of HD 179821 by Reddy \& Hrivnak (1999) and the assumption that when the star was losing mass as a red hypergiant, as much carbon and oxygen were
incorporated into CO as possible.  With $T_{mb}$ = 1.4 K (the average of the values reported by Josselin \& Lebre 2001 and Bujarrabal et al. 1992),
$V_{\infty}$ = 32 km s$^{-1}$ and $D$ = 6 kpc, then ${\dot M}$ = 3 ${\times}$ 10$^{-4}$ M$_{\odot}$ yr$^{-1}$. 
This inferred value of ${\dot M}$ is so large that the models by Kastner (1992) are not particularly accurate, but, as noted below, our CO maps are consistent with this result.  For a steady-state wind with $V_{\infty}$ = 32 km s$^{-1}$ and ${\dot M}$ = 3 ${\times}$ 10$^{-4}$ M$_{\odot}$ yr$^{-1}$,  $n(H_{2})$ = 1.4 ${\times}$ 10$^{4}$ $R_{17}^{-2}$ cm$^{-3}$ where $R_{17}$ denotes the distance from the star (10$^{17}$ cm). 

Josselin \& Lebre (2001) report recent single-beam observations of the CO emission in the  
$J$ = (1${\rightarrow}$0) and $J$ = (2${\rightarrow}$1) transitions.  Their
measured integrated intensities agree within 20\% with those found by Bujarrabal et al. (1992) for the $^{12}$CO emission, but they disagree by a factor of
3 for the $^{13}$CO line.  Using essentially the same integrated intensity for the $^{12}$CO line as we do, and assuming a distance of 1 kpc, Josselin \& Lebre (2001) estimate a mass loss rate of  2 ${\times}$ 10$^{-4}$ M$_{\odot}$ yr$^{-1}$.  Scaled to a distance of 1 kpc, we would estimate a mass loss rate
of 8 ${\times}$ 10$^{-6}$ M$_{\odot}$ yr$^{-1}$, a factor of 25 smaller than that found by Josselin \& Lebre (2001). There are three contributions to this difference.  First, Josselin \& Lebre (2001)  use a  formula very similar to equation (1) but with a  numerical coefficient that is a factor of 2 larger to estimate the mass loss rate from the CO  ($J$ = (1${\rightarrow}$0) 
line profile. Their equivalent to equation (1) is derived by Loup et al. (1993) based on a scaling procedure rather than  on detailed theoretical calculations as
employed by Kastner (1992).   Second,
we assume that [CO]/[H$_{2}$] = 10$^{-3}$ while Josselin \& Lebre (1993)
adopt [CO]/[H$_{2}$] = 5 ${\times}$ 10$^{-4}$, another factor of 2 in the estimate of the mass loss rate.  Third,
Josselin \& Lebre (1993) implicitly assume that the CO emission is spatially limited and does not fill the telescope beam.  This leads to the requirment that there must be more gas in a smaller volume and thus a factor of 6 increase in their estimate of  ${\dot M}$ over our inferred value.  In view of the theoretical calculations by
Mamom et al. (1988) and our  map, we see no need to make this
adjustment.  

With ${\dot M}$ $>$ 10$^{-5}$ M$_{\odot}$ yr$^{-1}$, the gas density is sufficiently high that the rotational levels of the CO molecule are mainly populated by collisions with H$_{2}$ (Kastner 1992).  Thus,
we can interpret our maps of the CO emission by comparing them  with   models for collisional excitation of CO which have been developed for interpreting observations of interstellar matter (Castets et al. 1990, Sakamoto 1996).  In the outer envelope around HD 179821 where $n(H_{2})$ $<$ 10$^{3}$ cm$^{-3}$, the CO is subthermally excited so  that the population in the J = 1 rotational level is lower than predicted by the Boltzmann weighting function.  In this case, a collisional excitation of CO into the J = 1 level is followed by radiative de-excitation so the  observed Rayleigh-Jeans brightness temperature, $T_{B}$, of the J = (1${\rightarrow}$0) CO emission depends  directly upon the collisional excitation rate.   Consequently, $T_{B}$ scales as n(H$_{2}$)$^{2}g(T_{K})$ where $g(T_{K})$ is a slow function of  the kinetic temperature, $T_{K}$. For the likely case that $T_{K}$ $>$ 20 K (see below), $g(T_{K})$ varies as $T_{K}^{0.25}$ (Castets et al. 1990).  

We now consider  the azimuthally-averaged CO
intensity at 4${\arcsec}$ offset (or $R_{17}$ = 3.6) from the star, a region where the density is probably low enough ($n(H_{2}$) ${\sim}$ 1100 cm$^{-3}$ from above) that this simple model for collisional excitation pertains, yet also a location where our data have a relatively high signal to noise.   At this offset of 4${\arcsec}$, $T_{B}(0)$ the brightness temperature in the channel at 0 km s$^{-1}$,  is 5.5 K or 160 mJy/beam (see Figure 4).  To reproduce this observation, we assume [CO]/[H$_{2}$] = 10$^{-3}$, $T_{K}$ = 60 K (see below) and  a spherically symmetric outflow.  In this case, for a sightline  whose impact parameter is $R_{impact}$,  $dV/dR$, the velocity gradient used in the calculations by Castets et al. (1990), is approximately given by $V_{\infty}/R_{impact}$.  With these parameters,  n(H$_{2}$) = 600 cm$^{-3}$ at $R$ = $R_{impact}$ (Castets et al. 1990).  This estimated density at $R_{17}$ = 3.6 is in reasonable agreement with the prediction from the model derived above with ${\dot M}$ = 3 ${\times}$ 10$^{-4}$ M$_{\odot}$ yr$^{-1}$ and $V_{\infty}$ = 32 km s$^{-1}$ that n(H$_{2}$) = 1100 cm$^{-3}$.  Therefore, within a factor of 2, it seems that an average
mass loss rate of 3 ${\times}$ 10$^{-4}$ M$_{\odot}$ yr$^{-1}$ can explain
both the observed CO integrated line profile (Bujarrabal et al. 1992) and the maps presented here.  

Additional evidence in favor of the model for
subthermal collisional excitation of the  CO is given by the angular variation of $T_{B}$. In the outer envelope of the nebula, the  model predicts that $T_{B}(0)$ scales as $n(R_{impact})^{2}R_{impact}$.  If 
 ${\dot M}$ and $V_{\infty}$ are constant, then $n(R_{impact})$ varies as $R_{impact}^{-2}$, and thus  $T_{B}(0)$ is predicted to  vary  as ${\phi}^{-3}$ where ${\phi}$ is the offset angle from the central star.  As shown in Figure 4, in the channel at 0 km s$^{-1}$, the CO intensity falls
from 270 mJy/beam (${\phi}$ = 3${\arcsec}$) to 40 mJy/beam  (${\phi}$ = 6${\arcsec}$).  This decrease of $T_{B}$ as ${\phi}^{-2.8}$ is in reasonable agreement with the expectation of the model for subthermal collisional excitation.  Furthermore, if $T_{B}$ varies as ${\phi}^{-3}$, then between an inner radius, ${\phi}_{in}$ and and outer radius, ${\phi}_{out}$, the total flux varies as (${\phi}_{out}$ - ${\phi}_{in}$).  With ${\phi}_{in}$ = 2${\arcsec}$, the approximate measured value of the inner radius of the CO ring, and ${\phi}_{out}$ = 6${\arcsec}$ and 11${\arcsec}$ for the OVRO interferometer and IRAM 30m telescope measurements,
respectively, then it is expected that F(OVRO)/F(IRAM) = 4/9 or 0.44 in
agreement with the measured values of 0.45 to 0.55 reported above.

With ${\dot M}$ = 3 ${\times}$ 10$^{-4}$ M$_{\odot}$ yr$^{-1}$, then ${\dot M}\,V_{\infty}$ equaled 6 ${\times}$ 10$^{28}$ g cm s$^{-1}$ when the star was losing a large amount of mass.  Currently,  $L_{*}/c$  equals 4 ${\times}$ 10$^{28}$ g cm s$^{-1}$.  This approximate agreement between ${\dot M}\,V_{\infty}$ and $L_{*}/c$ is  consistent with models for winds driven by radiation pressure
on dust (Lamers \& Cassinelli 1999), observations of other mass-losing red hypergiants (Jura \& Kleinmann 1990) and expectations for the pre-supernova evolution of massive stars (Heger et al. 1997).   Also,  this mass loss rate derived from the gas of 3 ${\times}$ 10$^{-4}$ M$_{\odot}$ yr$^{-1}$ compares with the value of 4 ${\times}$ 10$^{-4}$ M$_{\odot}$ yr$^{-1}$ derived by Jura \& Werner (1999) from the emission by the circumstellar dust  with an assumed gas to dust ratio of 100.  

The observed diameter of the CO emission of more than
 12${\arcsec}$, which is substantially larger than the  diameter of
the OH emission of ${\sim}$ 4-5${\arcsec}$ (Claussen 1993), can be understood in the usual  models for the photo-dissociation of molecules by ambient interstellar ultraviolet radiation as they flow out of the star (Glassgold 1996).  The OH molecule is mainly protected by dust and therefore can
be photodissociated at ${\sim}$2${\arcsec}$ from the star (Jura \& Werner 1999).  The CO is self-shielding, and
for a model with ${\dot M}$ = 3 ${\times}$ 10$^{-4}$ M$_{\odot}$ yr$^{-1}$ and $V_{\infty}$ = 30 km s$^{-1}$, the predicted fractional
abundance is  reduced as a result of  photodissociation by a factor of 0.5 at a radius of about 1.4 ${\times}$ 10$^{18}$ cm (or 16${\arcsec}$ for HD 179821) from the star (Mamon, Glassgold \& Huggins 1988).  Thus, in agreement with observations, the models predict the CO  persists to much greater distances from the star than does the OH.  Once the CO is photodissociated, the atomic carbon is rapidly photo-ionized
by the ambient interstellar ultraviolet.  As a consequence, little neutral
carbon is expected to be found, consistent with the upper limit for the C I emission reported by Knapp et al. (2000).

The observed circumstellar envelope around HD 179821 is consistent with that expected from theoretical models for the evolution of massive stars.
In the computations by Schaller et al. (1992), stars with initially 25 M$_{\odot}$ and 40 M$_{\odot}$ have maximum mass loss rates as red hypergiants of 3 ${\times}$ 10$^{-5}$ M$_{\odot}$ yr$^{-1}$ and 3 ${\times}$ 10$^{-4}$ M$_{\odot}$ yr$^{-1}$, respectively.  In the calculations by Chin \& Stothers (1990), the mass
loss rate for a star of 30 M$_{\odot}$ at times reaches 4 ${\times}$ 10$^{-4}$ M$_{\odot}$ yr$^{-1}$, the star  shrinks to approximately ${\sim}$10 M$_{\odot}$ and the duration of intense mass loss phase is  ${\sim}$8 ${\times}$ 10$^{4}$ yr.     

\subsection{Deviations from Spherical Symmetry}
While the spherically symmetric model can reproduce much of the data, it  fails
to explain all the observations. Consider the  apparent size
of the CO shell as a function of gas velocity.  
For a thin shell or radius, $R_{0}$, expanding at uniform velocity, $V_{\infty}$,  the 
 projected distance from the center of the nebula, ${\Delta}R$, at velocity offset from line center, ${\Delta}V$, is (Olofsson et al. 2000):
\begin{equation}
{\Delta}R/R_{0}\;=\;(1\,-\,[{\Delta}V/V_{\infty}]^{2})^{1/2}
\end{equation}
Figure 6 shows a comparison between the full width half maximum of the 
azimuthally-averaged CO intensity in the different velocity intervals. The model is a  reasonable first approximation to explain the data, but there may be deviations by a factor of 20\% in the outflow velocity from that predicted by
this model.

Jura \& Werner (1999) found that HD 179821 does not lie at the center of
the circumstellar ring, and a plausible interpretation of their data
is that the outflow speed varies by as much as ${\pm}$20\% from its average
value.  Also, as shown in  Figure 5, the profiles of the blue and red components of the circumstellar CO line at the center of the nebula disagree with each other, and therefore the outflow kinematics are not the same in the two directions.  Also, neither the blue nor the red component of the line profile
can be reproduced by a single outflow velocity.

To interpret the anisotropy in the CO map and to learn about the deviations from spherical symmetry, we model
the excitation of the CO molecule. First, we argue that the gas kinetic temperature is likely to be above 20 K.   
 Following Jura, Kahane \& Omont (1988) and Kastner (1992), 
the gas temperature is determined by a balance between  heating caused by the supersonic streaming of dust grains through the envelope and cooling resulting both from radiation and the adiabatic expansion of the matter.  In the outer portion of the circumstellar shell, where adiabatic cooling is more important than radiative cooling and where
${\dot M}\,V_{\infty}$ ${\approx}$ $L_{*}/c$, the gas temperature, $T_{K}$, is given by the expression:
\begin{equation} 
T_{K}\;{\approx}\;R^{-1}\,[{\sigma}_{grain}\,n_{grain}/n{_H}]\,Q^{3/2}\,(L_{*}/c)\,(4{\pi}k_{B})^{-1}
\end{equation}
where (${\sigma}_{grain}\,n_{grain}/n{_H}$) denotes the grain number density times the average geometric grain cross section divided by the number density of hydrogen nuclei, $Q$ is the ratio of the effective cross section for momentum transfer of a grain compared to its geometric cross section and $k_{B}$ is Boltzmann's constant.  Plausible but uncertain values of $Q$ = 0.01 and (${\sigma}_{grain}\,n_{grain}/n{_H}$) = 10$^{-21}$ cm$^{2}$ (Jura et al. 1988) yield $T_{K}$ = 220 $R_{17}^{-1}$ K.  

Since it is likely that $T_{K}$ $>$ 20 K in the regions which produce the CO emission, then  the inferred mass loss rate is relatively insensitive to the gas temperature and the unknown 
spatial variations in the grain properties.  Instead,
$T_{B}$ in a particular velocity interval at any offset angle varies
as the rate of collisional excitation or as the square of the local density, so  $T_{B}$ scales
as $({\dot M}/V_{\infty})^{2}$ (see equation 4 below).    Therefore, the observed factor of 2 variations in 
$T_{B}$ at offset angle of 4${\arcsec}$     
 are consistent with  $V_{\infty}$ and  ${\dot M}$  deviating in different directions  by
 as much as a factor of 1.4 from maximum to minimum.  According to Jura \& Werner (1999), a  possible explanation for these variations is the presence of strong
magnetic fields within the atmosphere of the mass-losing star (Soker 1998), but another model is a highly dipolar global convective flow (Jacobs, Porter \& Woodward 1999).  Magnetic fields sufficiently strong to affect the mass outflow morphology have been reported around the red hypergiant VX Sgr (Trigilio, Umana \& Cohen 1998). Current models for magnetized winds from red giants (Langer, Garcia-Segura \& Mac Low 1999, Garcia-Segura \& Lopez 2000) do not reproduce the asymmetry that we observe
around HD 179821; the observed circumstellar envelope around this star may provide useful constraints on
future calculations.     

\section{COMPARISON OF THE NEBULA AROUND HD 179821 WITH KEPLER'S SNR}
We now consider the possible fate of the nebula around HD 179821 when the star explodes as a supernova, and we compare it to Kepler's
remnant which is only 400 years old.  It has been previously suggested that the anisotropy in Kepler's remnant results from
a bow shock produced by a wind from the pre-supernova (Bandiera 1987).  Evidence for a pre-supernova wind is given by the  knots of optical nebulosity in the circumstellar remnant with proper motion speeds near 100 km s$^{-1}$ (Bandiera \& van den Bergh 1991). (This gas moving at ${\sim}$100 km s$^{-1}$ may have been partly accelerated by the supernova explosion and thus have been produced by a wind
with a speed near 30 km s$^{-1}$.)  
Here, instead of a model with a bow-shock, 
we suggest that the anisotropies in Kepler's supernova remnant can be
understood if the star exploded into an intrinsically anisotropic circumstellar wind similar to
that currently found around HD 179821.

There is a qualitative similarity between the 11.7 ${\mu}$m and CO maps of the shell around HD 179821  with the 21 cm continuum VLA map (Reynoso \& Goss 1999) and the ROSAT  map (Hughes 1999) of soft X-rays ($E$ $<$ 2 keV) of Kepler's supernova remnant.  Both objects show very approximately circular symmetry, but with one side being markedly brighter than the opposite.  However, there 
is also a quantitative difference between the two structures.  In HD 179821, the contrast between the integrated intensity of CO to the north and south
is about a factor of 1.4.  The contrast of the peak intensity between the northern and southern hemispheres in the 
ROSAT image of Kepler's supernova remnant is approximately a factor of 7. Even though the X-ray images of Kepler's supernova remnant display a much greater contrast in intensity than does the CO image of HD 179821, below
we suggest that the nebula-like that around HD 179821 may evolve into
an object like Kepler's supernova remnant.  

A full hydrodynamical calculation of a spherically symmetric supernova explosion into an anisotropic 
pre-supernova wind is beyond the scope of this paper.  Here, we consider a simplified  version of the interstellar bubble models presented by Weaver et al. (1978).
Let the supernova begin as a spherically symmetric explosion of energy $E_{0}$ and propagate without
significant radiative losses into a detached envelope whose density, 
${\rho}(R)$, is
described as
\begin{equation}
{\rho}\;=\;{\dot M}/(4\,{\pi}\,V_{\infty}\,R^{2})
\end{equation}
for $R$ ${\geq}$ $R_{detached}$ and ${\rho}$ = 0 for $R$ $<$ $R_{detached}$
where $R_{detached}$ denotes the inner boundary of the detached circumstellar shell.  
With the very approximate assumption that most of the mass in the observed remnant
is swept-up material, $M_{sw}$, then:
\begin{equation}
1/2\,M_{sw}\,{\dot R}^{2}\;{\approx}\;{\epsilon}_{kin}\,E_{0}
\end{equation}
where ${\epsilon}_{kin}$ denotes the fraction of the supernova's explosive energy that goes into kinetic energy (instead of  heating the gas).   We may take ${\epsilon}_{kin}$ = 0.4 (Weaver et al. 1978).  In the limit that $R$ $>$ $R_{detached}$, then:
\begin{equation}
M_{sw}\;{\approx}\;{\dot M}\,R(t)/V_{\infty}
\end{equation}
The solution to equations (5) and (6) is:
\begin{equation}
R\;=\;(9\,{\epsilon}_{kin}\,E_{0}\,V_{\infty})^{1/3}(2{\dot M})^{-1/3}\,t^{2/3}
\end{equation}
If $E_{0}$ = 10$^{51}$ erg, $t$ = 400 yr and $R$ = 8 ${\times}$ 10$^{18}$ cm (Hughes 1999), then  equation (7)
yields ${\dot M}/V_{\infty}$ = 5.5 ${\times}$ 10$^{14}$ g cm$^{-1}$.  If $V_{\infty}$ = 32 km s$^{-1}$, then ${\dot M}$ was 3 ${\times}$ 10$^{-5}$ M$_{\odot}$ yr$^{-1}$.   These parameters for the presupernova wind are moderately similar to those inferred for the envelope around HD 179821, which therefore may be following an evolutionary path parallel to that experienced by the star that produced Kepler's supernova.   

We now
estimate the angular variation of the X-ray emission that might occur when the current shell around HD 179821 is shocked by a symmetric supernova explosion.  
When viewed tangentially, the intensity of the X-ray emission, $I_{X}$, can
be written as:
\begin{equation}
I_{X}\;{\propto}\;f(T)\,{\rho}(R_{impact})^{2}\,R_{impact}
\end{equation}
where ${\rho}(R_{impact}$) is the density at the impact parameter, $R_{impact}$, and $f(T)$ is a complicated function of the gas temperature.
At the expansion speed of Kepler's supernova remnant  of
1500-2000 km s$^{-1}$ (Blair, Long \& Vancura 1991),  the post-shock gas temperature is ${\sim}$ 3 ${\times}$ 10$^{7}$ K, and the emissivity of the gas near  1 keV is relatively insensitive to the gas temperature (Raymond \& Smith 1977).  We
therefore ignore variations in $f(T)$, and
from equations (4) and (8) write:
\begin{equation}
I_{X}\;{\propto}\;({\dot M}/V_{\infty})^{3}\,E_{0}^{-1}\,
t^{-2/3}
\end{equation}
Since our radio data indicate that  $({\dot M}/V_{\infty})$ varies by a factor of 1.4, the predicted variation  of the X-ray intensity is about a factor of 
3, which is somewhat smaller than the observed factor of 7 contrast in the X-ray emission.

 Chin \& Stothers (1990) found that a star of initially 30 M$_{\odot}$ on the main sequence shrinks to 11.3 M$_{\odot}$ through the red hypergiant phase and then, after 76,000 yr,  explodes as
a supernovae.  If this model applies to HD 179821, then during the next 76,000 yr, the shell expelled at 32 km s$^{-1}$ will expand to an inner radius of 7.7 ${\times}$ 10$^{18}$ cm which is approximately equal to the  radius of Kepler's SNR of 8 ${\times}$ 10$^{18}$ cm (Hughes 1999).

\section{THE AFTERMATH OF THE SUPERNOVA}
Above, we have considered the effect of a symmetric explosion impacting upon an asymmetric envelope and the resulting supernova remnant.  However, it is also imaginable that just as the outer material is ejected anisotropically during the slow wind phase, the inner material might similarly be ejected anisotropically during the supernova explosion.  Such an asymmetric explosion  could explain why some pulsars have space velocities near 700 km s$^{-1}$  (Cordes \& Chernoff 1998).  If the total momentum loss, ${\vec p}$, is expressed as the sum of Legendre polynomials, and
if this expansion is simplified  to just the first two
terms,  then:
\begin{equation}
{\vec p}\;=\;[P_{0}\;+\;P_{1}\,cos({\theta})]\,[1/(4{\pi})]\;{\hat r}
\end{equation}
where ${\theta}$ is measured relative to the presumed axis of symmetry and
${\hat r}$ is a unit vector in the radial direction.  
The total $z$-component of the momentum carried away by the ejecta, $p_{z}$, is
\begin{equation}
p_{z}\;=\;(2{\pi})\,{\int}^{\pi}_{0}\,({\vec p}{\cdot}{\hat z})\,\,sin({\theta})\,d{\theta}\;=\;(1/3)\,P_{1}
\end{equation}
The magnitude of the total momentum carried away by the ejecta, $p_{t}$, is
\begin{equation}
p_{t}\;=\;(2{\pi})\,{\int}^{\pi}_{0}\,({\vec p}{\cdot}{\hat r})\,\,sin({\theta})\,d{\theta}\;=\;P_{0}
\end{equation}
If the supernova explodes with energy, $E_{0}$, and the material moves
outward with velocity, $V_{explode}$, then $p_{t}$ = 2$E_{0}/V_{explode}$.  If there is a neutron star of mass, $M_{ns}$, left behind after the supernova explosion, then the magnitude of the velocity kick, $V_{kick}$, acquired by the neutron star can be inferred from the assumption that the momentum carried by the neutron star is balanced by $p_{z}$, the momentum carried by the remnant, so that:
\begin{equation}
V_{kick}\;=\;(1/3)\,P_{1}/M_{ns}\;=\;(2/3)\,(P_{1}/P_{0})\,E_{0}/(M_{ns}
\,V_{explode})
\end{equation}
Above, we have argued that the anisotropic outflow from HD 179821 is consistent with variations of ${\dot M}$ and $V_{\infty}$ by 20\% from the mean and therefore we adopt $P_{1}/P_{0}$ = 0.2.  With standard values of $E_{0}$ = 10$^{51}$ erg, $M_{ns}$ = 1.4 M$_{\odot}$ and $V_{explode}$ = 10$^{4}$ km s$^{-1}$, then equation (12) yields $V_{kick}$ = 500 km s$^{-1}$.  Given that these parameters may  vary by as much as a factor of 2 from the assumed values, it is possible
that kicks as large as 700 km s$^{-1}$ may result from supernova explosions if
they are as asymmetric as the outflow from HD 179821.  

It is possible that supernova explosions produce gamma-ray bursts.  If so, then  the late-time light curves of these
systems are in part determined by the propagation of the explosion into the pre-existing circumstellar nebula (Esin \& Blandford 2000).     When HD 179821 explodes as a supernova, perhaps in 8 ${\times}$ 10$^{4}$  yr, the inner radius of its current dust shell may have expanded to 2.6 pc.  Esin \& Blandford (2000) propose that
supernovae possess dust shells at a distance in the range 0.1-1 pc to explain the 20-30 day delay in the dust echoes from gamma-ray bursts.  The detached shell around HD 179821 may be  representative
of the environments where dust echoes from gamma-ray bursts are produced.  
\section{CONCLUSIONS}
We have obtained high resolution J = (1${\rightarrow}$0) CO observations of the circumstellar shell around HD 179821, a star which will probably become a supernova in the
next 10$^{5}$ yr.  We make the following points:
\begin{itemize}
\item{To reproduce the CO map and spectra, it seems that until about 1600 years ago, the star was a red hypergiant with an 
average mass loss rate near 3 ${\times}$ 10$^{-4}$ M$_{\odot}$ yr$^{-1}$ at an average outflow speed of 32 km s$^{-1}$.}
\item{The north-south asymmetry in the  CO emission resembles that seen at 11.7 ${\mu}$m.  The CO data suggest that ${\dot M}/V_{\infty}$ varies by  as much as a factor of 1.4 from minimum to maximum.  There appear to be deviations by as much as factor of 1.2 in the expansion velocity of the gas.}
\item{The asymmetry seen in the infrared and molecular emission around HD 179821 resembles that
 seen in the radio and X-ray emission from Kepler's supernova remnant. Even if a supernova explosion is spherically symmetric, it might propagate into an intrinsically asymmetric circumstellar nebulae and thus even young remnants can appear asymmetric.}
\item{If HD 179821 explodes as an asymmetric supernova with the momentum loss varying by as much as 1.2 in different directions, consistent with an extrapolation of our observations of its outer circumstellar envelope to the inner core, then the pulsar that is created may have a space motion approaching 700 km s$^{-1}$.}
\item{The detached dust shell around HD 179821 may be representative of
the environments which produce dust echoes 20-30 days after a gamma-ray burst.}
\end{itemize}
We have had useful conversations with A. Kusenko and S. Phinney as well as email
correspondence with R. Chevalier and E. Josselin.  
This work has been partly supported both at UCLA by NASA and at the Jet Propulsion Laboratory, California Institute of Technology, under an agreement with NASA.  The Owens Valley Radio Observatory Millimeter Array is supported in part by the National Science Foundation grant AST 96-13717.

\newpage
\begin{center}
{\bf FIGURE CAPTIONS}
\end{center}
Fig. 1.  Plot of the LSR Velocity vs. the Outflow Velocity derived from CO data 
with squares for the  oxygen-rich high mass loss rate (${\dot M}$ ${\geq}$ 
10$^{-6}$ M$_{\odot}$ yr$^{-1}$) AGB stars within 1 kpc of the Sun from Jura \& 
Kleinmann (1989).  The crosses for HD 179821 and the well known hypergiant 
IRC+10420 are also shown.  The data are consistent with the hypothesis that a 
high outflow velocity is correlated with a high luminosity, and that both HD 
179821 and IRC+10420 are distant, high luminosity objects.  
\\
\\
Fig. 2. The circumstellar dust and gas shells around HD 179821. The upper panel 
shows the J = 1-0 $^{12}$CO  intensity superimposed upon the 11.7 ${\mu}$m map 
obtained by Jura \& Werner (1999); the lower map shows the CO data by itself. 
The spatial  resolution of each map (1{\farcs}4$ \times$ 
2{\farcs}0 for CO and 0{\farcs}4$ \times$ 0{\farcs}4 for 11.7 ${\mu}$m) is shown. The CO intensities represent  the integrated line 
intensities ${\int}T\,dv$ over a velocity spread of 65 km s$^{-1}$ centered at 
V$_{lsr}$ = 100 km s$^{-1}$.    The first contour is at 142.4 K km s$^{-1}$; the contours levels are spaced by 71.2 K km s$^{-1}$ (1.8${\sigma}$).  The star's 
position is denoted as a red cross.  
The color-coding in the lower panel is a redundant indicator of the intensity of the emission. In both panels the highest intensity is represented by red and the lowest by blue.  The image of HD 179821 in the 11.7 ${\mu}$m image is noncircular because of an imperfect chop-nod motion of the Keck telescope
during the time that the data were acquired.   
\\
\\
Fig. 3. Channel maps of the circumstellar  J = 1-0 $^{12}$CO emission from the 
shell around HD 179821 shown with 5.35 km s$^{-1}$ velocity spacing.  The star's position is denoted as a  cross.  The first contour intensity in each map is at 
80 mJy beam$^{-1}$ (2.6 K  brightness temperature); the contours levels are 
spaced by 80 mJy beam$^{-1}$ (2.3${\sigma}$). 
\\
\\
Fig. 4.  Azimuthally-averaged intensity of the CO emission (mJy/beam) in the 
channel map at line center (0 km s$^{-1}$) and the channel map at +27 km 
s$^{-1}$ velocity offset as a function of angular offset (${\arcsec}$) from the 
central star.
\\
\\
Fig. 5. The CO spectrum (mJy/beam vs. velocity) at the center of the circumstellar envelope. 
\\
\\
Fig. 6.  Plot of the FWHM of the CO emission (${\arcsec}$) for different velocity channels compared to the expectation from equation (2).  For the model fitting,
we take, the half width radius = 3${\farcs}$95 and $V_{\infty}$ = 32 km s$^{-1}$.  The error bars in the FWHM of the CO emission are taken as 0${\farcs}$2 or 10\% of the beamsize.   
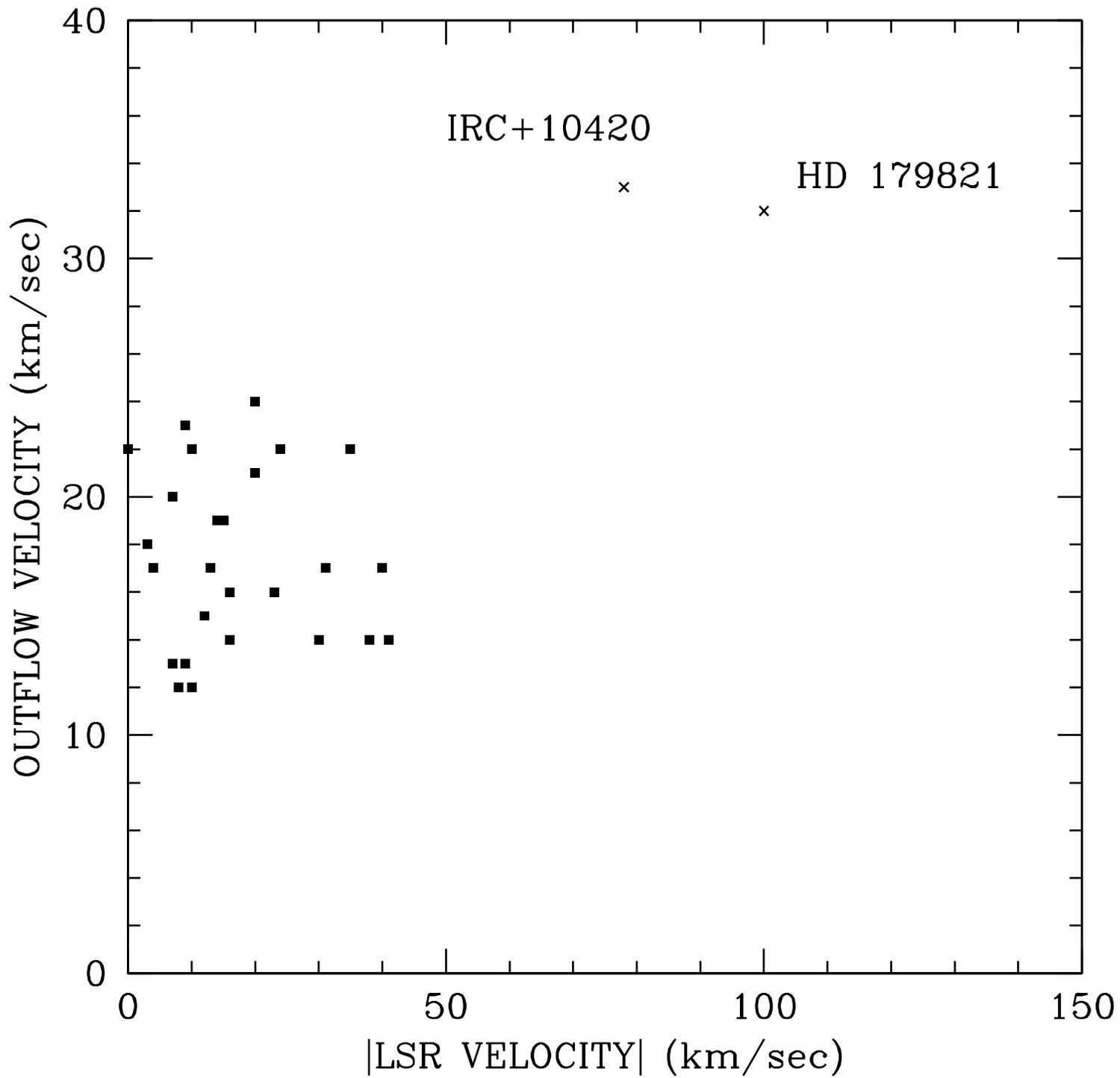
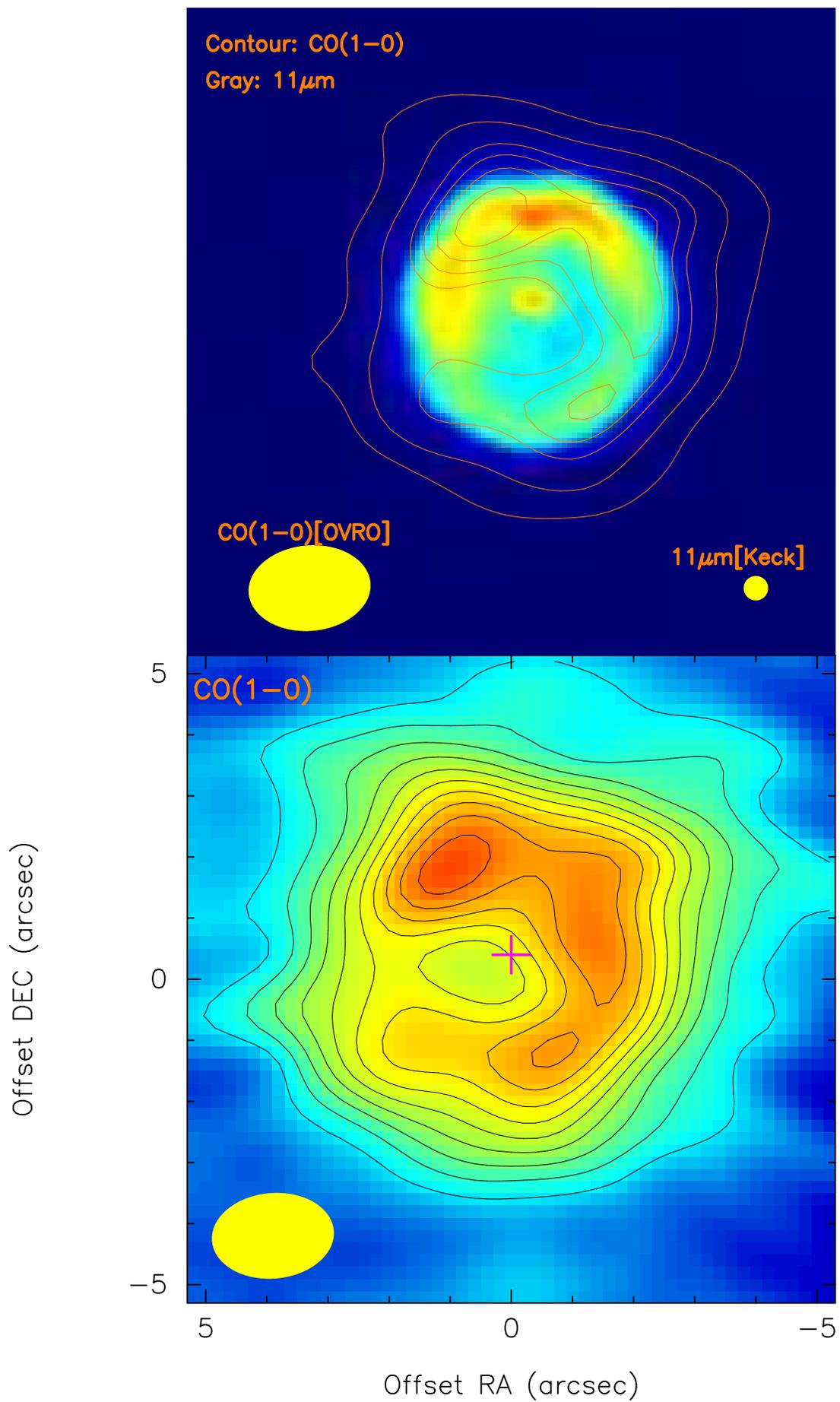
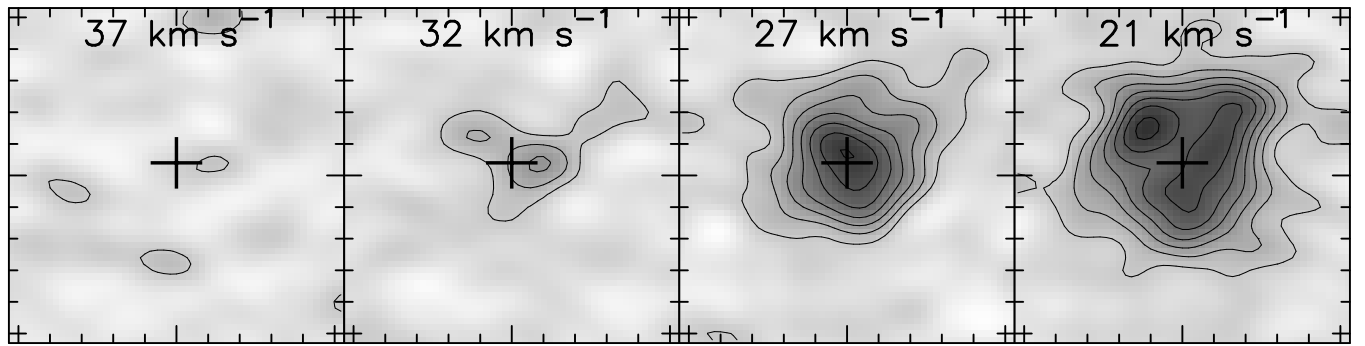
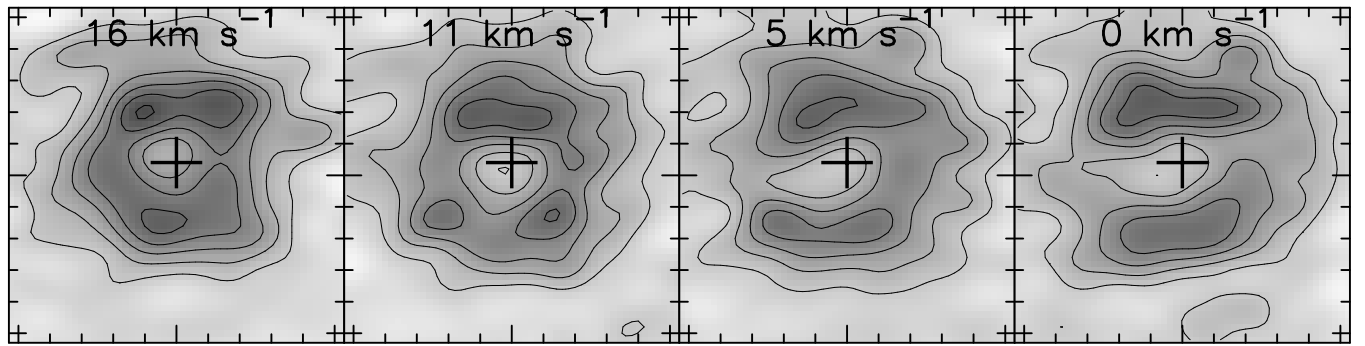
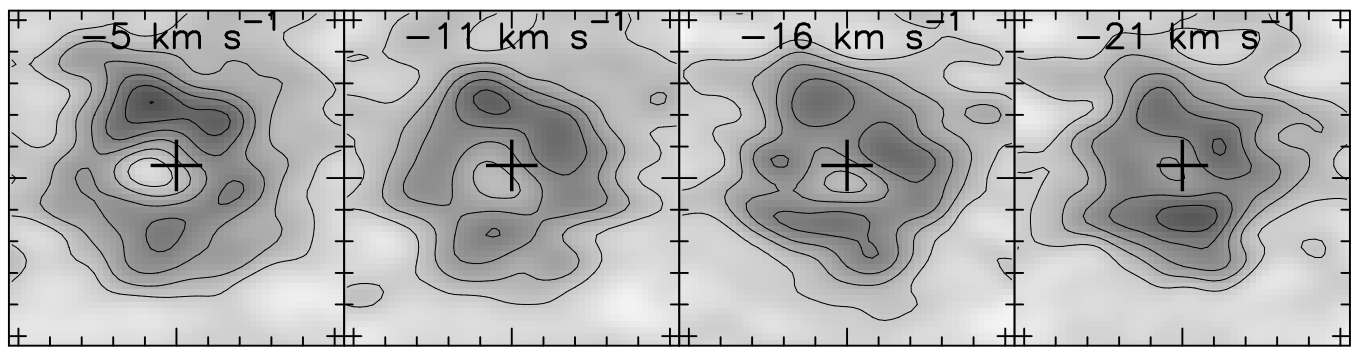
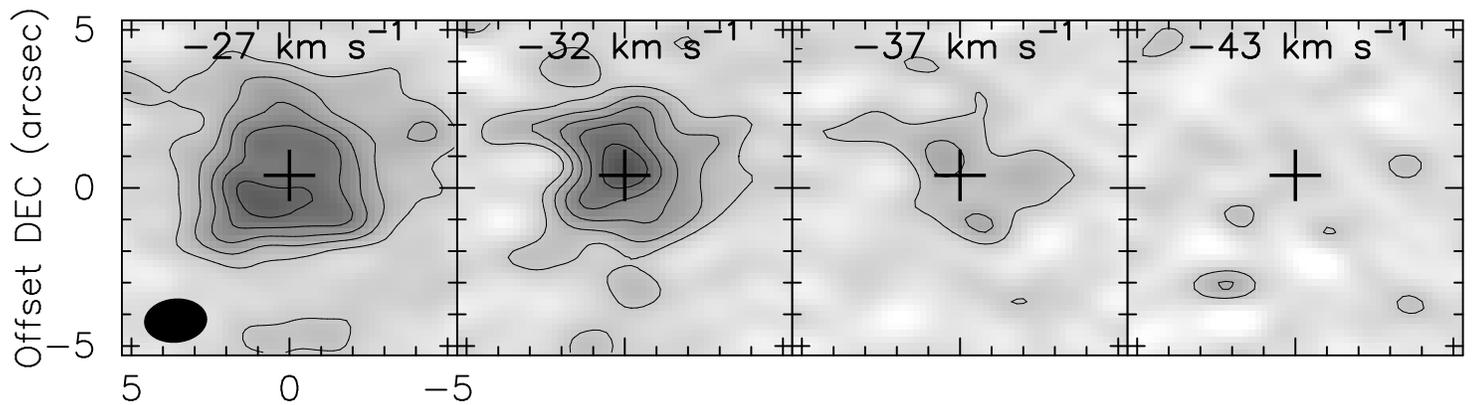
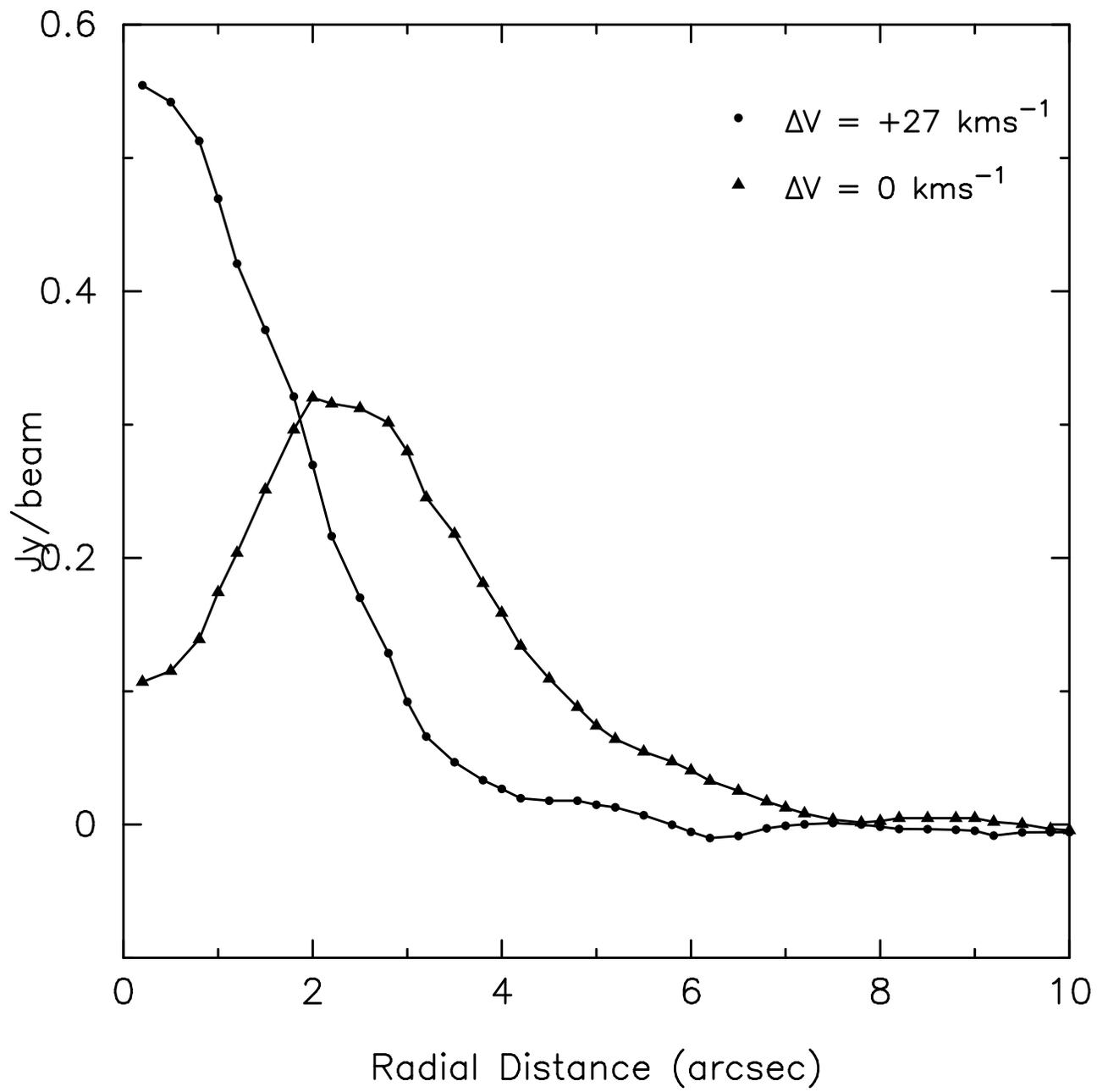
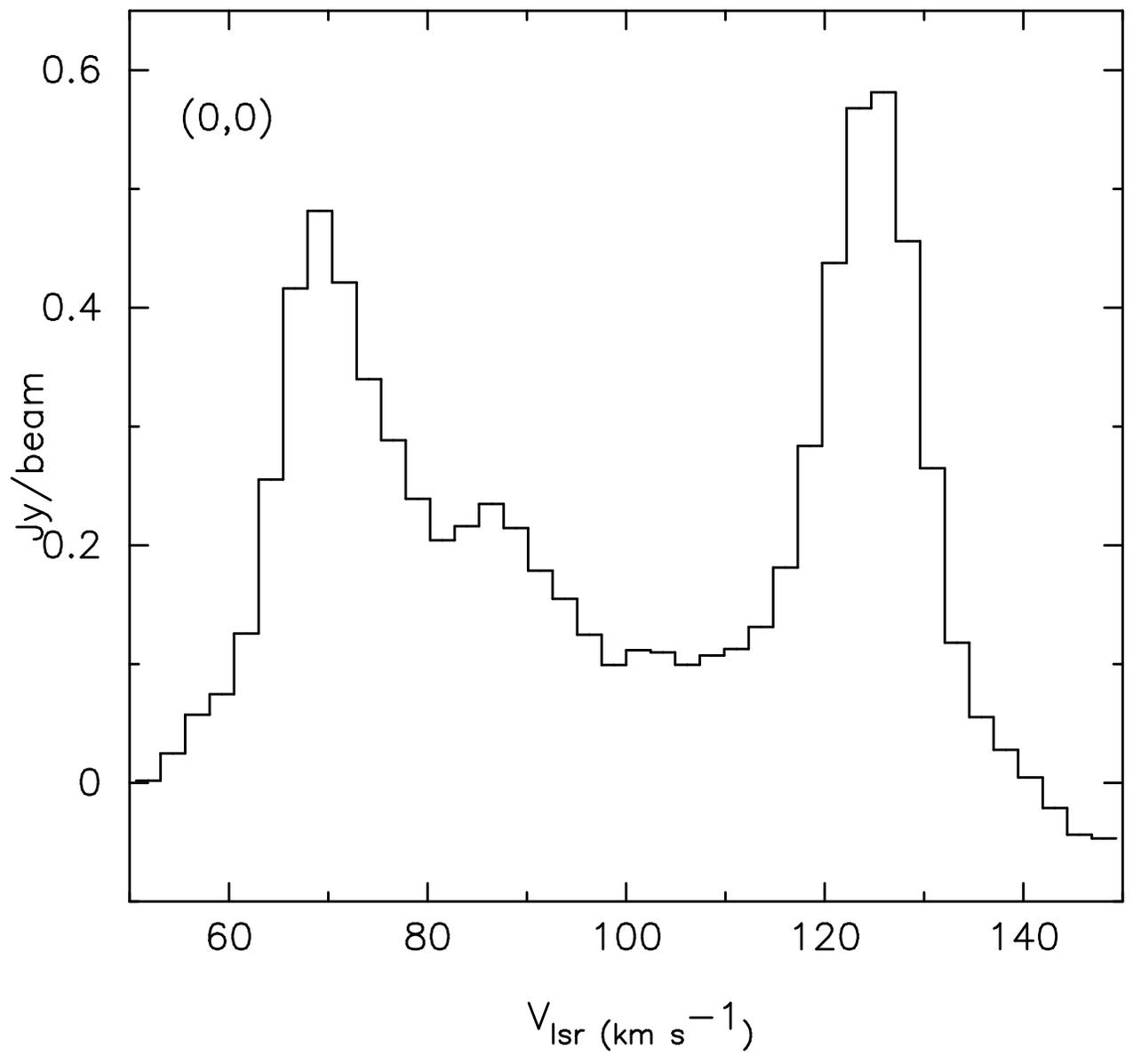
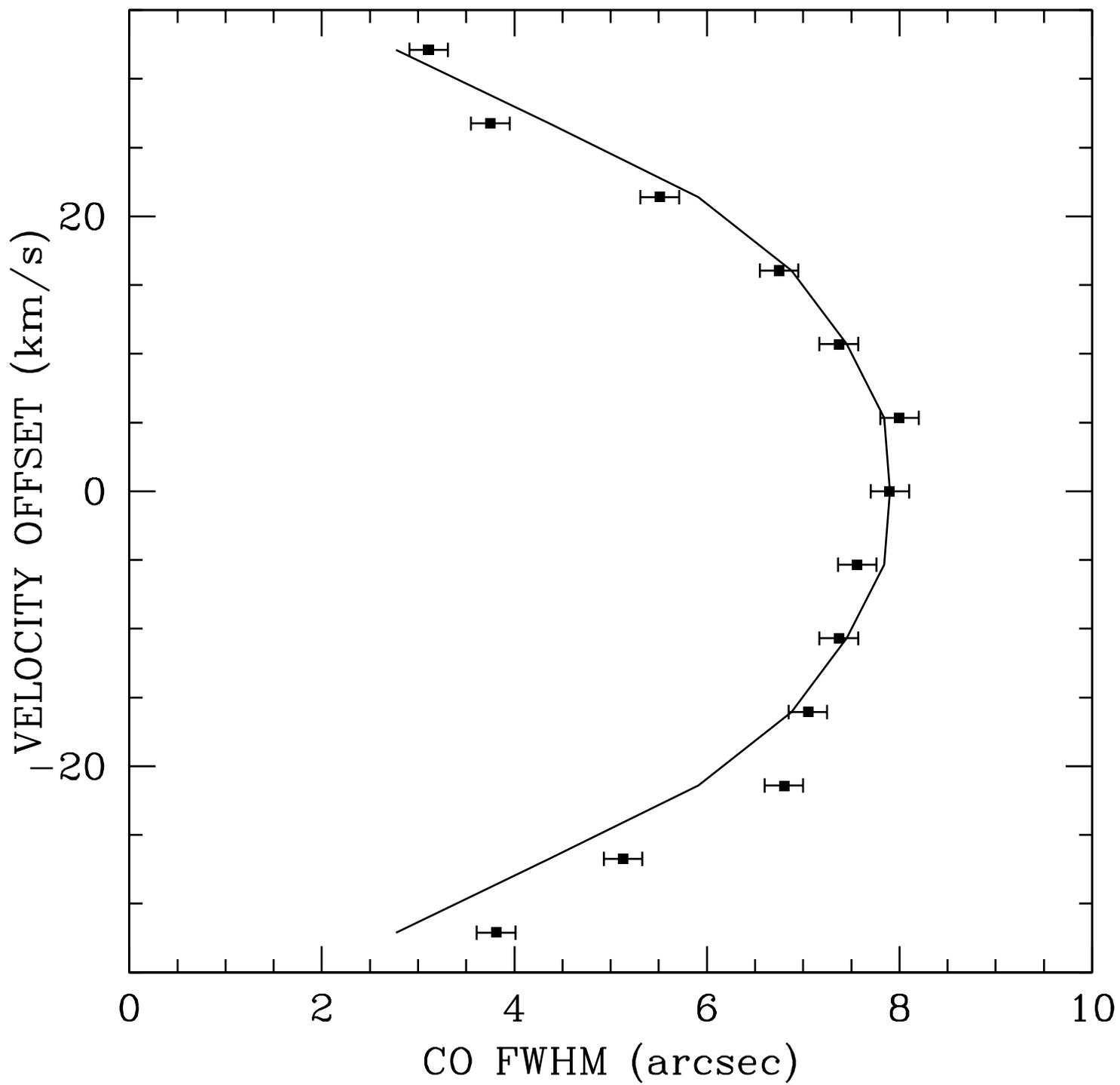
\end{document}